\documentclass[a4paper, 12pt]{article}

\usepackage{longtable}
\usepackage{setspace}
\usepackage{graphicx}

\usepackage[ansinew]{inputenc}
\usepackage[english]{babel}
\usepackage[round,longnamesfirst]{natbib}
\usepackage{color}
\usepackage{graphicx,setspace}
\usepackage{amsmath}
\usepackage{amssymb}
\usepackage{subfig}
\usepackage[usenames,dvipsnames]{xcolor}
\usepackage{epstopdf}

\newcommand{\Diag}[1]{\mbox{\rm Diag}\!\left(#1\right)} 
\newcommand{\Normal}[1]{\mathcal{N}\left(#1\right)}
\newcommand{\Betadis}[1]{\mathcal{B}\left(#1\right)}

\newcommand{\Gammainv}[1]{\mathcal{G}^{-1} \left(#1\right)}

\newcommand{\Normult}[2]{\mathcal{N} _{#1}\left(#2\right)}

\newcommand{\alphav}{\boldsymbol{\alpha}}
\newcommand{\gammav}{\boldsymbol{\gamma}}
\newcommand{\betav}{\boldsymbol{\beta}}
\newcommand{\deltav}{\boldsymbol{\delta}}
\newcommand{\epsilonv}{\boldsymbol{\epsilon}}
\newcommand{\kappav}{\boldsymbol{\kappa}}
\newcommand{\lambdav}{\boldsymbol{\lambda}}
\newcommand{\muv}{\boldsymbol{\mu}}
\newcommand{\thetav}{\boldsymbol{\theta}}
\newcommand{\omegav}{\boldsymbol{\omega}}

\newcommand{\sigmav}{\boldsymbol{\sigma}}

\newcommand{\varepsilonv}{\boldsymbol{\varepsilon}}
\newcommand{\zetav}{\boldsymbol{\zeta}}

\newcommand{\Lambdav}{\boldsymbol{\Lambda}}

\newcommand{\Sigmav}{\boldsymbol{\Sigma}}
\newcommand{\Thetav}{\boldsymbol{\Theta}}
\newcommand{\Psiv}{\boldsymbol{\Psi}}

\newcommand{\fv}{\mathbf{f}}

\newcommand{\vv}{\mathbf{v}}
\newcommand{\wv}{\mathbf{w}}
\newcommand{\xv}{\mathbf{x}}
\newcommand{\yv}{\mathbf{y}}
\newcommand{\zerov}{\mathbf{0}}
\newcommand{\Fv}{\mathbf{F}}
\newcommand{\Iv}{\mathbf{I}}

\newcommand{\Sv}{\mathbf{S}}
\newcommand{\Wv}{\mathbf{W}}

\DeclareMathOperator{\V}{V}
\DeclareMathOperator{\E}{E}
\DeclareMathOperator{\diag}{diag}
\DeclareMathOperator{\Cov}{Cov}

\begin{document}
	\onehalfspacing
	
	\begin{center}
\Large{Factor-augmented  Bayesian  
treatment  effects models  for panel outcomes \\[5pt]
	\vspace{0.5cm} }
\normalsize
\textsc{Helga Wagner}\footnote{Department of Applied Statistics, Johannes Kepler University Linz, Altenberger Stra\ss e 69, 4040 Linz, Austria.
	E-mail: Helga.Wagner@jku.at  {\bf Corresponding author.}}\\[5pt]
\textsc{Sylvia Fr\"uhwirth-Schnatter}\footnote{Department of Finance, Accounting and  Statistics, Vienna University of Economics and Business,
	Geb\"aude D4, 4. Stock, Welthandelsplatz 1, 1020 Vienna, Austria. Phone: +43-1-313 36-5581. Fax: +43-1-313 36-774.
	E-mail: Sylvia.Fruehwirth-Schnatter@wu.ac.at}\\[5pt]
\textsc{Liana Jacobi}\footnote{Department of Economics, FBE Building, Level 4, 111 Barry Street, The University of Melbourne, VIC 3010, Australia.
	E-mail: ljacobi@unimelb.edu.au}
\\[5pt]

\vspace{0.5cm}

\today \\[5pt]
\vspace{0.5cm} \emph{keywords}: endogeneity, bifactor model; switching regresson model; shared factor model,  dynamic treatment effects  \vspace{0.5cm}

\textsc{Abstract}
\end{center}

\begin{quote}   We
	propose  a new,  flexible  model  for inference of  the  effect  of a binary treatment on a continuous outcome observed over subsequent time periods. The model allows to seperate   association due to endogeneity  of treatment selection   from additional longitudinal association of the
outcomes and hence  unbiased estimation of dynamic treatment effects.
 We  investigate the performance  of the proposed method on simulated data and
employ it to  reanalyse data on the longitudinal effects of a long maternity leave
  on mothers' earnings after their return to the labour market.
\end{quote}



\section{Introduction} \label{sec:intro}
Identification and estimation of treatment effects  is an important issue in many  fields, e.g. to evaluate
 the effectiveness of social programs, government policies or medical
interventions. As each subject is observed  only either  under  control conditions or under treatment,   the outcome difference which would allow straightforward estimation of treatment effects is not available for any particular subject.
Additionally, for data from observational studies,  endogeneity of treatment selection can cause unobserved confounding and bias of  treatment effects estimates if not adequately accounted for.

Bayesian  approaches to inference on treatment effects  rely on  specifying a joint model of treatment selection and
 the two potential outcomes, under control  conditions and under treatment,   of which
  only one is observed for each subject.   To estimate  the  effect  of a binary treatment on a 
  continuous outcome observed over subsequent time periods two models,
the switching regression model \citep{chi-jac:mod} and the shared factor model
\citep{car-etal:est},  have been  suggested  so far.

Both approaches rely on a binary regression model for selection into treatment and two multivariate regression models for the outcome sequences under control and under treatment, however they differ with respect to  modeling  the dependence
across  these regression models: Whereas \cite{car-etal:est} model the association between treatment selection and both
 potential outcome
 sequences via  shared latent factors, \cite{chi-jac:mod}  specify only two marginal models for selection into treatment and one sequence of potential outcomes but leave the joint distribution of the two potential outcomes sequences  unspecified.

 Investigating both models  in detail,   \citet{jac-etal:bay_tre}
 show that both frameworks  
 impose  restrictions on the joint correlation structure of
 treatment selection
and the two outcomes sequences that 
can result in biased treatment effects estimates if the
 assumptions on the  correlation structure
 of the model used for data analysis are violated  in  the data generating process.
 
To increase flexibility  in the dependence structure of treatment selection and potential
outcomes  we propose  in the present paper a factor-augmented treatment effect model which 
extends the factor structure of the joint distribution to a  bi-factor model.
 The bi-factor model  was  introduced in  \cite{hol-swi:bif} and recently
 gained popularity in item response analyses, see e.g. \cite{rei:red}. Its basic assumption is  that the covariance  structure of multiple responses  can be modelled by orthogonal factors where one common (or general) factor is  shared by all responses
and one or more further group (or specific) factors model the additional correlation among clusters of responses. This  is  attractive for jointly modelling of treatment selection and the two potential outcomes sequences as it  allows to  model
association due to endogeneity  of treatment selection
as well as additional longitudinal association of the
outcomes sequences:  the general factor shared by the binary selection and both potential outcomes sequences accounts  for unobserved confounding  whereas outcome specific factors   allow to  model the  additional longitudinal association that cannot be attributed to the  unobserved confounders.

The paper is structured as follows. Section~\ref{sec:trtmod} discusses Bayesian treatment
 effects models for panel outcomes and  reviews the switching regression and the shared factor model.
  Section~\ref{sec:bifac} introduces the 
  factor-augmented treatment effect model,
 discusses identification issues and  describes posterior inference using MCMC methods.
 In Section~\ref{sec:sim},  the  flexibility  of the factor-augmented treatment effect model is illustrated on simulated data.
 Section~\ref{sec:appl} provides   a reanalysis of  longitudinal effects of a long maternity leave
  on mothers' earnings after their return to the labour market and  Section~\ref{sec:conclusion} concludes.

 \section{Bayesian modelling of panel treatment effects } \label{sec:trtmod}
 Assessing the effect of a treatment on an outcome of interest requires a comparison  of this outcome
 under two  conditions: with and without  treatment ( control conditions). As typically each subject is
 observed only under either treatment or control conditions modelling of treatment effects  relies on
 the potential outcomes framework \citep{rub:est}, which allows  to define treatment effects based on
 models for the outcome under treatment as well as under control conditions. However,
  inference on treatment effects from observational data is demanding as in addition to the fundamental problem  that only one potential outcome is observed for
   each subject,  treatment is not randomized,  but self selected and hence might be associated to
   the outcome.

 To take  endogeneity of treatment selection into account, Bayesian approaches to modelling
 treatment effects  rely on
 specifying a joint model for treatment selection and the  potential outcomes,
   often in the spirit of Roy's switching regression model \citep{roy:som,lee:uni}.
   For longitudinally  observed outcomes, two approaches have been suggested sofar:
   \cite{chi-jac:mod} specify two  models for selection into treatment and one potential outcomes
   sequence respectively, whereas \cite{car-etal:est} specify a joint model
 for selection into treatment and the two potential outcomes models.
 \cite{jac-etal:bay_tre} use both approaches to analyse the effects of a longer maternity leave on
 the earnings of mothers.

While the switching regression model as well as   the shared factor model rely on a probit
model for treatment selection and two multivariate normal regression models for the potential
 outcomes sequences, they differ with respect to
modelling their joint distributions.
To discuss these differences in more detail, we introduce  the
marginal  models for latent utilities and the potential outcomes
sequences in Section~\ref{ssec:meanmod} and describe modelling of
  the  dependence structure in both approaches in Section~\ref{ssec:dependendence}.

\subsection{Marginal models for treatment  selection and  outcomes sequences} \label{ssec:meanmod}

 Let  $x_i, i=1,\dots, n $  denote  the treatment status of each
 subject $i$   for   $i=1,\dots, n$. Treatment selection depends on  covariates
 (\emph{selection on observables}) via a  probit model for $x_i$, which can be  specified  in terms of
a latent Gaussian random variable $x_i^*$   as
\begin{align} x_i^* & =\vv_i \alphav +\varepsilon_{xi}, \qquad
 & \varepsilon_{xi} \sim \Normal{0,\sigma^2_x}, \label{eq:util0}\\
	x_i & =I_{\{x_i^*>0\}} \label{eq:trt0},
\end{align}
where $\vv_i$ denotes the row  vector of covariates  and
$\alphav$  their effect  on treatment selection. Note that different from the usual specification of a
probit model, (\ref{eq:util0})  leaves  the variance of the error term unspecified,
 whereas usually the error variance  of the latent utility is fixed to 1, $\sigma^2_x=1$,
  as regression  effects $\alphav$ are only identified up to a scale factor.  However, in
  factor models  where the error term  $\varepsilon_{xi}$ is modelled  by a latent factor plus
  an idiosyncratic 
  error  it is more convenient to fix the variance of the  idiosyncratic 
  error to one.

 The selection model given in  equations  (\ref{eq:util0}) and (\ref{eq:trt0})  is combined with a model
  for  the potential outcomes for subject $i$ at time points $t=1,\dots, T$ which we denote
  by $y_{0,it}$  and  $y_{1,it}$  for the outcome under control conditions and
 treatment, respectively. The potential outcomes  are  modelled as
 \begin{align}
 \label{yj chib re0t}
 y_{0,it} & = \eta_{0,it}(\wv_{it})+\varepsilon_{0,it}, \qquad   &\varepsilon_{0,it}
 \sim \Normal{0,\sigma^2_{0t}},\\
 \label{yj chib re1t}
 y_{1,it} & =\eta_{1,it}(\wv_{it})+ \varepsilon_{1,it}, \qquad&  \varepsilon_{1,it}
 \sim \Normal{0,\sigma^2_{1t}},
 \end{align}
 with   structural means $\eta_{j,it}(\wv_{it})$  for $j=0,1$ depending  a  row
 vector of covariates
  $\mathbf{w}_{it}$:
 \begin{align}
 	\eta_{0,it} (\wv_{it}) & =\mu_t+ \mathbf{w}_{it} \boldsymbol{\gamma}
 	\label{linpred_0t}\\
 	\eta_{1,it} (\wv_{it})& =(\mu_t + \kappa_t)+  \mathbf{w}_{it}
 (\boldsymbol{\gamma}+\boldsymbol{\theta} ). \label{linpred_1t}
 \end{align}
Here, $\mu_t$ and $\mu_t +\kappa_t$ are the intercepts and
 $\boldsymbol{\gamma}$ and  $\boldsymbol{\gamma}+ \boldsymbol{\theta}$ the vectors of covariate effects
   under,  respectively, control conditions and under treatment.
 With this specification, the average treatment effect
 of a subject with covariate values $\mathbf{w}_{it}$  in panel period $t$ results
as
\begin{align*}
E({y}_{1,it}- y_{0,it}| 
\mathbf{w}_{it}) = \kappa_t+ \mathbf{w}_{it}\thetav.
\end{align*}
However,  the  observed outcome $y_{it}|(x_i=j)$
conditional on knowing $x_i=j$ is equal to $y_{0,it}$, if $x_i=0$, and
equal to $y_{1,it}$, if $x_i=1$.
Or,  in terms of the latent utility $ x_i^*$ introduced
 in equation (\ref{eq:util0}):
$$y_{it}  | x_i^* =\begin{cases} y_{0,it}, \quad &  \text{for } x_i^*< 0 ,\\
	y_{1,it}, \quad & \text{for }  x_i^*\ge 0 .\end{cases}$$
In randomized studies, treatment
selection $x_i$ is independent from  the observed outcome. Hence,
$y_{it}|(x_i=j)$ has the same distribution as $y_{j,it}$,
which allows straightforward estimation of the average treatment effects from the observed outcomes.
This is, however, not the case
in observational studies where subjects  choose treatment based
on their expectations on the outcomes and therefore explicit modelling of
the  association between treatment selection and observed outcome is necessary.
 We return to this issue in Section~\ref{ssec:dependendence}.

In the following, we will denote the vectors of potential outcomes by
$\yv_{ji}=({y}_{j,i1}, \dots,  {y}_{j,iT})$ for $j=0,1$
 and  by $\yv_{i}|(x_i=j)=\big({y}_{i1}|(x_i=j), \dots,  {y}_{iT}|(x_i=j)\big)$  the vector of observed outcomes for subject $i$.

\subsection{Modelling the dependence structure}\label{ssec:dependendence}

As noted above, endogeneity of treatment selection  (\emph{selection on unobservables}) can be taken
into account  by allowing for correlation of treatment selection and the potential outcomes.
This approach is followed both by  the switching regression as well as the
shared factor model.
 However, the impossibility to observe both outcomes $\yv_{0i}$ and $\yv_{1i}$
 for the same subject $i$ makes it impossible to observe the joint distribution
    of the errors  $(\varepsilonv_{0i},\varepsilonv_{1i})'$ and the two models
    differ with respect to specifying the  error distribution.

In the   switching regression model, the joint distribution
    of $(\varepsilonv_{0i},\varepsilonv_{1i})$ is left unspecified and only
   the two joint  $(T+1)$-variate distributions of   the
  latent utility error $\varepsilon_{xi}$ and  the errors 
  $\varepsilonv_{ji}=(\varepsilon_{j,i1}, \dots, \varepsilon_{j,iT})'$
  in each outcome equation,
   i.e.~the marginal distributions of $(\varepsilon_{xi}, \varepsilonv_{0i})$ and
   $(\varepsilon_{xi}, \varepsilonv_{1i})$, are
    specified  as multivariate normal distributions.

 In contrast, in the shared factor model   the joint  $(2T+1)$-dimensional  distribution
 of all error terms ($\varepsilon_{xi}, \varepsilonv_{0i}, \varepsilonv_{1i})$
  is  modelled in terms of  latent factors and  independent idiosyncratic errors.
 \citet{ car-etal:est}  specify a multi-factor model and exploit additional measurements from psychological tests to identify factors and factor loadings.
 Such additional measurements  are  not available in  the analysis  of \cite{jac-etal:bay_tre} who  therefore use a simpler factor structure
 with only one subject specific random factor  that accounts for  within subject dependence as well as  endogeneity. They specify
 the error terms as
 \begin{align}\varepsilon_{xi}&= \lambda_x f_i +\epsilon_{xi}, \qquad  &&\epsilon_{xi} \sim \Normal{0,1}, \label{errx}\\
 \varepsilonv_{ji}& = \lambdav_j f_i + \epsilonv_{ji},\qquad
 &&\epsilonv_{ji} \sim \Normal{\zerov,\Sv_j=\diag(\sigma^2_{j1},\dots,\sigma^2_{jT})}, \quad  j=0,1, \label{errj}
 \end{align}
 where  $f_i \sim \Normal{0,1}$ is an unobserved subject specific factor, $\lambda_x$ is its loading on  the
 latent utility and $\lambdav_j=(\lambda_{j1},\dots, \lambda_{jT})'$, $j=0,1$ denote the vectors of factor loadings  for the potential outcomes.
  $\epsilon_{xi}$ and $\epsilonv_{ji}$, $j=0,1$ are  the idiosyncratic errors of the latent utility and the potential outcome vectors, respectively. Both, factor loadings
  $\lambda_{jt}$  as well as  the  variances  $\sigma^2_{jt}$ of the idiosyncratic  errors are  allowed to vary over time.
 The joint covariance matrix of the vector  $\varepsilonv_i=(\varepsilon_{xi},\varepsilonv_{0i}', \varepsilonv_{1i}')'$ is  then given as
 $$\Cov(\varepsilonv_i)=\Sigmav=\begin{pmatrix} \sigma_{x}^2 & \sigmav_{x0}' & \sigmav_{x1}'\\
 \sigmav_{x0} & \Sigmav_0 & \Sigmav_{01}\\
 \sigmav_{x1} & \Sigmav_{01} & \Sigmav_{1}\end{pmatrix}=
 \begin{pmatrix}1+\lambda_x^2 & \lambda_x \lambdav_0' & \lambda_x \lambdav_1'\\
 \lambda_x \lambdav_0 &  \lambdav_0 \lambdav_0'+\Sv_0 &  \lambdav_0 \lambdav_1'\\
 \lambda_x \lambdav_1 &    \lambdav_0 \lambdav_1' & \lambdav_1 \lambdav_1'+\Sv_1 \end{pmatrix}.
 $$
 Hence, for fixed covariates, $\sigmav_{xj}=\Cov(x_i^*, \yv_ {ji})$ denotes  the vector of
  covariances between the latent utility $x_i^*$ and the potential outcome vector $\yv_{ji}$,
  and $\Sigmav_{j}=\Cov(\yv_{ji})$ is the covariance matrix of the potential outcome vector $\yv_{ji}$
   for $j=0,1$. Most importantly,
    the covariance matrix  of  the two potential outcome vectors $\yv_{0i}$ and $\yv_{1i}$,
    $\Sigmav_{01}=\Cov(\yv_{0i},\yv_{1i})$,  is modelled explicitly.
 The assumption that  the  latent factor $f_i$ is shared by the latent utility and all potential outcomes implies that
 the  vectors of time-varying factor loadings $\lambdav_j$, $j=0,1$ determine not only
  $\Sigmav_{j}$ (and thus the correlation within each potential outcome vector) and
 $\sigmav_{xj}= \lambda_x \lambdav_j$ (and thus the correlation between
  latent utility and each potential outcome vector),
 but also $\Sigmav_{01}$  and thus the correlation between the potential outcome vectors.
 Note, however, that this assumption is not testable from empirical data where only one potential outcome is observed for each subject.

 In  contrast, in the switching regression model  a  latent factor is  assumed to model only
 the correlations within one potential outcomes vector
 $$ \varepsilonv_{ji} = \lambdav_j f_{ji} + \epsilonv_{ji},
 \qquad  \epsilonv_{ji} \sim \Normal{\zerov,\Sv_j}, \quad  j=0,1, $$
 where  (as in the shared factor model)   $\Sv_j=\diag(\sigma^2_{j1},\dots,\sigma^2_{jT})$
 is the variance matrix of the idiosyncratic errors. The latent factors  $f_{0i}$ and $f_{1i}$
  are assumed to have  standard normal  marginal distributions, $f_{ji} \sim \Normal{0,1}$,
  but  no assumption is made  on their
 joint distribution. While the error term of the latent utility,
  $\varepsilon_{xi}=\epsilon_{xi}$,   is independent of the factors, it
 is  allowed to be  correlated with the idiosyncratic errors  $\epsilonv_{ji}$ of each
 potential outcome equation, to capture selection on unobservables.
 Hence,  the joint $(T+1)$-variate distributions of the idiosyncratic errors
 $(\epsilon_{xi},\epsilonv_{ji}')$
 are given as
 \begin{equation}(\epsilon_{xi},\epsilonv_{ji}')'  \sim \Normal{\zerov,\begin{pmatrix} 1 & \omegav_j' \\
 	\omegav_j &\Sv_j\end{pmatrix}},  \quad  j=0,1, \label{eq:covsr} \end{equation}
 where   the vector  $\omegav_j$ of covariances  between the latent utility and
 potential outcome vector $\yv_{ji}$ is  completely unstructured.

 Both models have drawbacks. The shared factor model relies on the untestable assumption that the
  latent factor is shared by the latent utility and both potential outcomes. Hence, as the factor
   loadings determine all correlations in the multivariate normal distribution of the errors
    $\varepsilonv_{ji}$, this correlation structure is not fully flexible.
    On the other hand, in the switching regression model
     no joint model for the error terms  is specified and
      the variance of the outcome difference is not available.  Additionally,  the assumption that
      each latent factor  affects only the corresponding potential outcomes vector,
        but not the latent utility,
 implies that,  conditional on the latent utility error $\epsilon_{xi}$,
the  idiosyncratic errors
 $\epsilonv_{ji}$ in the potential outcome model are negatively correlated over time,
  since equation (\ref{eq:covsr}) implies
  following  covariance matrix of $\epsilonv_{ji}$ given $\epsilon_{xi}$:
 $$\Cov(\epsilonv_{ji}|\epsilon_{xi})=\Sv_j - \omegav_j\omegav_j'.$$

 None of the models encompasses the other,
  but the  positive semi-definiteness of the specified covariance matrices can result
  in restrictions on their elements in one model  which cannot be recovered under the other. Thus,
    as shown in a simulation study in \citet{jac-etal:bay_tre},
    treatment effects can be biased,  when data generated  under the shared factor model
     are analysed using the switching regression model and vice versa.

\section{A factor-augmented treatment effects model} \label{sec:bifac}

In this section we propose a factor-augmented model for modelling the joint distribution
of the latent utility and the two potential outcomes sequences to allow
for a flexible dependence structure where the correlation within an outcomes
 sequence is disentangled into correlation due to confounding and additional longitudinal  correlation.
We introduce the model in Section~\ref{sec:specification}
  and discuss identification issues  in  Section \ref{ssec:identification}. Section  \ref{ssec:pri} describes specification of the prior distributions and Section  \ref{ssec:mcmc} outlines posterior inference.

\subsection{Model specification} \label{sec:specification}

 We consider the model specified in Section \ref{ssec:meanmod}, with a probit model for treatment
 selection given in equations (\ref{eq:util0}) and (\ref{eq:trt0}) and
 the model for the potential outcome sequences
 $\yv_{0i}$ and $\yv_{1i}$ given as
 \begin{align}
 \yv_{0i} & =\muv+\Wv_i \thetav+\varepsilonv_{0i}, \label{eq:panout0}\\
 \yv_{1i} &=\muv+ \kappav +\Wv_i ( \thetav+\gammav)+ \varepsilonv_{1i}, \label{eq:panout1}
\end{align} 	
where $\Wv_{i}$  is the corresponding matrix of covariate values with  rows
 $\mathbf{w}_{i1}, \ldots, \mathbf{w}_{iT} $ and  $\muv$ and $\kappav$ are the vectors
$\muv=(\mu_1,\dots, \mu_T)$ and  $\kappav=(\kappa_{1}, \dots, \kappa_T)$.


 To  achieve more flexibility in modelling  the association of treatment selection and the   potential outcomes  sequences,    we assume that  all  dependencies  in  the error vector
 $\boldsymbol{\varepsilon}_i=(\varepsilon_{xi}, \varepsilonv_{0i}',\varepsilonv_{1i}')'$
 are captured by  three  subject specific latent factors:
 one common  factor   $f_{ci}$  which is  shared by the error terms of the
latent utility $x^*_i$  and both potential outcome vectors $\mathbf{y}_{0i}$ and $\mathbf{y}_{1i}$  and two specific factors
$f_{0i}$ and $f_{1i}$  which affect only the error vectors of the potential outcome $\varepsilonv_{0i}$
and $\varepsilonv_{1i}$ respectively.

The  common factor thus  accounts for unobserved confounding whereas
 the two    outcome  specific factors $f_{0i}$ and $f_{1i}$
  capture  the additional longitudinal association in the   outcome vectors
that cannot be attributed to unobserved confounders. The joint model for the error terms is thus
 specified as
 \begin{align}
 \varepsilon_{xi} & =\lambda_x f_{ci }+\epsilon_{xi}, \qquad  & \epsilon_{xi} \sim \mathcal{N}(0,1)\label{m:errx},\\
  \boldsymbol{\varepsilon}_{0i} & =\boldsymbol{\lambda}_0 f_{ci }+ \boldsymbol{\boldsymbol{\zeta}}_0  f_{0 i }+
  \boldsymbol{\epsilon}_{0i},  \qquad   & \epsilon_{0,it}\sim \mathcal{N}(0,\sigma^2_{0t}),\label{m:err0}\\
  \boldsymbol{\varepsilon}_{1i} & =\boldsymbol{\lambda}_1 f_{ci } + \boldsymbol{\boldsymbol{\zeta}}_1 f_{1 i }+
  \boldsymbol{\epsilon}_{1i},
  \qquad   & \epsilon_{1,it}\sim \mathcal{N}(0,\sigma^2_{1t}) \label{m:err1},
 \end{align}
 where the  factors $f_{ci }$, $f_{0 i }$ and $f_{1 i }$ are  assumed to be
   independent  standard normals. Hence,  the factor loadings  $\lambda_x$, $\boldsymbol{\lambda}_j$ and $\boldsymbol{\zeta}_j$,   $j=0,1$
  determine the joint variance-covariance matrix of all error  terms.

  Assumptions on how factors are related to outcomes
 simplify the structure of the  factor loadings matrix. First,
   treatment selection and both  outcome panels depend  on the common factor
   $f_{ci}$ with loadings $\lambda_x$ in the latent utility and $\lambdav_j$, $j=1,2$
  	for the two potential outcome.
  	Second, the  potential outcomes vector $\yv_{ji}$ depends  only on one
  specific factor $f_{ji}$
  	with loadings $\zetav_j$.

In matrix form the factor model for the errors is given as
\begin{equation}
	 \varepsilonv_i= \Lambdav \fv_i +\epsilonv_i, \qquad  \Lambdav= \begin{pmatrix} \lambda_x & 0 & 0\\
	\lambdav_0 & \zetav_0 &\zerov_T\\
	\lambdav_1 & \zerov_T & \zetav_1 \end{pmatrix},
\end{equation}
where $\fv_i=(f_{ci},f_{0i},f_{1i})'$ is the vector of factors for subject $i$
and $\epsilonv_i=(\epsilon_{xi},\epsilonv_{0i}',\epsilonv_{1i}')'$ is the vector of idiosyncratic
 errors. Hence, the joint $(2T+1)$-variate distribution of $\varepsilonv_i$
 is multivariate normal,
  $\varepsilonv_i \sim \Normal{\zerov, \Sigmav}, $
  with    variance covariance matrix  given as
  $$\Sigmav= \begin{pmatrix}  \sigma_{x}^2 & \sigmav_{x0}' & \sigmav_{x1}'\\
  	\sigmav_{x0} & \Sigmav_0 & \Sigmav_{01}\\
  	\sigmav_{x1} & \Sigmav_{01} & \Sigmav_{1}\end{pmatrix}=
  \begin{pmatrix}1+\lambda_x^2 & \lambda_x \lambdav_0' & \lambda_x \lambdav_1'\\
  	\lambda_x \lambdav_0 &  \lambdav_0 \lambdav_0'+  \zetav_0 \zetav_0'+\Sv_0 &  \lambdav_0 \lambdav_1'+\zetav_0 \zetav_1'\\
  	\lambda_x \lambdav_1 &    \lambdav_0 \lambdav_1'+\zetav_0\zetav_1' & \lambdav_1 \lambdav_1'+ \zetav_1 \zetav_1'+ \Sv_1 \end{pmatrix}.
  $$

 Though an  extension  to multiple independent outcome specific factors $\fv_{ji}$
 is straightforward conceptually, to achieve identification of the factor loadings from the
 observed data  the  dimension of outcome specific factors
 is restricted by the number of available panel observations $T$. We will return to this issue in Section \ref{ssec:identification} but note here that in contrast to \cite{car-etal:est}
 we do not assume that additional measurements are available from which the latent factors can be identified.

  Already the simple factor-augmented 
  model  specified above   avoids drawbacks of both
  the shared factor and the switching regression model. As correlation across panel
   outcomes is not  attributed solely  to  the common factor,
  it  is more  flexible than  the  shared factor model
   which is  recovered as that special case where $\boldsymbol{\zeta}_1=\boldsymbol{\zeta}_2=\mathbf{0}$.
     Without any assumption on the joint distribution of the specific factors
  $f_{0i}$ and $f_{1i}$ the model is a switching regression model with the advantage that
   conditional on the
  latent factors the errors of latent utility and each potential outcome are independent.


\subsection{Identification}\label{ssec:identification}

 An important issue in factor models is their identification,   which according  to \cite{and-rub:sta}
  is a two-step procedure
 where the first step is  identification of the variance contribution attributable to the latent
 factors, i.e.   $\Lambdav\Lambdav'$  and the second step is identification of $\Lambdav$, i.e.
  solving the rotational identification problem.

  The data structure in the factor augmented  treatment model proposed above, however, differs  considerably
  from the  standard factor model  which assumes multivariate normal observations:
   only the binary treatment variable $x_i$ and the observed outcome sequence,
    which  is a truncated version of  one of the two potential outcomes sequences are
    observed for each subject. However, due to the general triangular structure of the factor
    loadings matrix rotational identification is not an issue for the bi-factor model,
    see \cite{fru-lop:spa}.

  In  the probit model, identification of regression effects is feasible only up to  the standard
   error $\sigma_x$ of the latent utility and hence  only the standardized effects
    $$\tilde\alphav=\frac{\alphav}{\sigma_x}=\frac{\alphav}{\sqrt{1+\lambda^2_x}}$$
    in model (\ref{eq:util0}) are identified.  Fixing $\sigma_x=1$ would in principle
     be possible, however require to restrict the range of the factor loading
     $\lambda_x$ to $(-1,1)$.

  The observed data never provide information on the association between the two potential outcome
  vectors $\yv_{0i}$ and $\yv_{1i}$.
   Due to endogeneity of treatment selection the distribution of the observed outcome sequence $\yv_i$
    is not the  marginal distribution of
  $\yv_{ji}$, but the conditional distribution truncated by the respective range of the
  latent utility. It is  given as
  \begin{equation*}
  	p(\yv_{ji}|x_i=j)=\begin{cases}
  		\frac{1}{1-\Phi(\tilde  \mu_{x,i})} \int_{-\infty}^0 p(\yv_{0i},x_i^*)d x_i^*,
   & \quad j=0,\\
  		\frac{1}{\Phi(\tilde  \mu_{x,i})} \int_0^\infty p(\yv_{1i},x_i^*)d x_i^*,
   & \quad j=1,
  		\end {cases}
  	\end{equation*}
  	where $\tilde  \mu_{x,i}= \vv_i \tilde \alphav$ is the mean of the
   standardized latent utility
  $x_i^*/\sigma_x$.

 Identification of all parameters that are not identified from the probit model, i.e. 
 the regression effects in the  potential outcomes models, the factor loadings and the variances of the
   idiosyncratic errors has to  be accomplished from these two conditional distributions.
   We will discuss necessary conditions  for identification of the model parameters from the first and
   second moments of these two conditional distributions.

The  expectation of the observed outcomes sequence $\yv_i|x_i=j$ is given as:
\begin{align*}
\E(\yv_i|x_i=0)& =\E(\yv_{0i}|x_i=0)=  \muv+ \Wv_i\gammav   -
\frac{\sigmav_{0x}}{\sigma_x} \frac{\phi(\vv_i \tilde \alphav) }{1-\Phi(\vv_i\tilde \alphav)}, \\
\E(\yv_i|x_i=1) & =\E(\yv_{1i}|x_i=1)=  \muv+ \kappav+ \Wv_i(\gammav+\thetav)+
 \frac{\sigmav_{1x}}{\sigma_x}\frac{\phi(\vv_i\tilde\alphav) }{\Phi(\vv_i \tilde \alphav)},
\end{align*}
see Appendix \ref{app:obsmoments} for details.  As the quantities $c_{i0}(\vv_i
\tilde \alphav)=- \frac{\phi(\vv_i\tilde \alphav) }{1-\Phi(\vv_i\tilde \alphav)}$
 and $c_{i1} (\vv_i \tilde \alphav )=\frac{\phi(\vv_i\tilde\alphav) }{\Phi(\vv_i\tilde \alphav)}$ are
identified from the probit model, identification
of the parameters  $\muv, \kappav,\gammav, \thetav$ and
 $\tilde \sigmav_{jx}=\frac{\sigmav_{jx}}{\sigma_x}$, for $j=0,1$ is feasible
 from these equations,
 if the design matrix in the corresponding regression model is of full rank.
$\frac{\sigmav_{0x}}{\sigma_x}$ and $\frac{\sigmav_{0x}}{\sigma_x}$ yield
$2T$ equations for the $2T+1$ factor loadings  $\lambdav$ of the common factor, leaving at least one of these factor loadings unidentfied.
The conditional covariance matrices $\V(\yv_i |x_i=0)$ and $\V(\yv_i |x_i=1)$,
given as
\begin{align*}\V(\yv_i |x_i=0) & =\Sigmav_{0}
	- c_{i0}(\vv_i  \tilde \alphav)\Big( \vv_i \tilde \alphav
+ c_{i0}(\vv_i \tilde \alphav)\Big) {\tilde \sigmav_{x0}}'{\tilde \sigmav_{x0}},
	\\
	\V(\yv_i |x_i=1) &=\Sigmav_{1} - c_{i1}( \vv_i
\tilde \alphav)\Big(\vv_i \tilde \alphav  + c_{i1}(\vv_i \tilde \alphav)\Big)
  {\tilde \sigmav_{x1}}'{\tilde \sigmav_{x1}},
\end{align*}
have $\frac{T(T+1)}{2}$ free elements each,
from which one element of $\lambdav$,  the factor loadings $\zetav_0$ and $\zetav_1$
and the variances $\sigmav^2_0$ and $\sigmav^2_1$ have to be identified.
   Thus, a necessary condition for identification is that
$$T(T+1)\ge 4T+1,$$
and hence that the observed panel outcomes are at least of length $T \ge 4$.
 Generally, as a necessary condition for  identification of the parameters in a model with $r>1$
  outcome specific factors is that
$$ T(T+1)  \ge  2(r+1)T +1,$$
the outcome panels have to be at least of length $T=6 $ and $T=8$ to identify the loadings of $r=2$ or $r=3$ outcome
specific factors, respectively.
 Identification of the elements of $\lambdav_{0}$,   $\lambdav_{1}$, $\zetav_0$ and $\zetav_1$, however,
 requires also that enough factor loadings are different from $0$, see
 \citet{and-rub:sta,con-etal:bay_exp, fru-lop:spa}.


\subsection{Prior distributions}   \label{ssec:pri}

To complete the  Bayesian model specification,
 prior distributions are assigned to all model parameters.
We write the model for the observed outcome vector of subject $i$ compactly as
$$y_{it}|(x_i=j) = \wv_{j,it}\betav +\varepsilon_{j,it},$$
 where  $\betav=(\muv, \kappav,\gammav, \thetav)$ comprises all
  regression parameters in both outcome models
  and $\wv_{j,it}$ denotes the corresponding covariates at panel time $t$.
 Thus the  factor-augmented (FA) treatment model is  given as
 \begin{align*}
 	x_i^*& = \vv_i \alphav+ \lambda_x f_i +\epsilon_{xi},  & \epsilon_{x i}
  \sim & \Normal{0,1},\\
 	\yv_{i}|(x_i=j) &= \Wv_{ji} \betav+\lambdav_j f_{ci} + \zetav_j f_{ji} + \epsilonv_{ji},
  & \epsilonv_{ji} \sim & \Normult{T}{\zerov,\mathbf{S}_j}, 
 \end{align*}
where the $t$th row of $\Wv_{ji}$ is equal to $\wv_{j,it}$.

We assume that the regression parameters $\alphav$, $\betav$,
the factor loadings $\lambda_x, \lambdav_0,\lambdav_1$, $\zetav_0$, $\zetav_1$ and the
variances of the idiosyncratic errors $\mathbf{S}_0, \mathbf{S}_1$
 are independent apriori.

Following \cite{jac-etal:bay_tre},  we perform variable selection in the
 selection as well as the outcome model to avoid overspecification.
 We assume prior independence of all coefficients in  $\alphav$ and $\betav$ and
 specify  normal priors $\Normal{0,V_\alpha}$
  and $\Normal{0,V_\beta}$, respectively,
 for the coefficients in $\alphav$ and $\betav$ not subject to selection.
 In  our application, these are the intercept in the selection equation and
  $\mu_1$ in the outcome equation and we use $V_\alpha=5$ and  $V_\beta=10^4$.

 For all coefficients in $\alphav$ and $\betav$ subject to selection, we
  employ spike and slab priors with a Dirac spike at 0.
  A spike and slab prior
  is a mixture of a component
 concentrated at zero, the spike, and a comparably flat component, the slab.
  A Dirac spike and slab prior has a Dirac spike at zero.
  For each coefficient   in $\betav$ subject to selection, $\beta_\ell$,  this prior is
    specified hierarchically as     $$p(\beta_\ell)= \delta^{\beta}_\ell p(\beta_\ell|\delta^{\beta}_\ell=1)
     +(1-\delta^{\beta}_\ell)      I_{0}(\beta_\ell),$$
     depending on a binary variable     $\delta^{\beta}_\ell$
  with prior inclusion probability $p(\delta^{\beta}_\ell=1)=\pi_{\beta}$ and
  $\pi_{\beta} \sim \Betadis{a,b}$. A similar prior is introduced for each coefficient
     in $\alphav$ subject to selection, $\alpha_\ell$, involving a binary indicator
    $\delta^{\alpha}_\ell $ with prior inclusion probability
    $p(\delta^{\alpha}_\ell=1)=\pi_{\alpha}$ and
  $\pi_{\alpha} \sim \Betadis{a,b}$.  In  our application, we use normal slabs with zero mean and variance  5
 and uniform priors $\Betadis{1,1}$ on the inclusion probabilities
    $\pi_\alpha$ and $\pi_\beta$.

 All indicator variables  are subsumed in the vectors
  $\deltav^{\alpha}$ and $\deltav^{\beta}$, respectively, and estimated along with
  the model parameters during MCMC estimation (see Section~\ref{ssec:mcmc}).
  Note that estimation of indicators $\delta^{\beta}_\ell$
  corresponding to coefficients in the parameter $\gammav$
  identifies relevant predictors for the outcome equation under control.
   Most importantly, estimation of indicators $\delta^{\beta}_\ell$ corresponding to
    coefficients in $\thetav$ leads to the identification of
  heterogeneous treatment  effects.

Finally, we assume that all elements in the  factor loading
 vectors, $\lambda_x, \lambdav_0,\lambdav_1$, $\zetav_0$, $\zetav_1$,  a priori are
 independent  standard normal $\Normal{0,1}$.
 Also the variances $\sigma^2_{jt}$ of the idiosyncratic errors are assumed to be independent
 and assigned a  $\Gammainv{s_{0j},S_{0j}}$ prior.
 We use  $s_{0j}=S_{0j}=2.5$ in our application.

\subsection{Posterior inference} \label{ssec:mcmc}

Posterior inference can be accomplished
by Markov chain Monte Carlo (MCMC) methods, extending \cite{jac-etal:bay_tre}.

As 
the idiosyncrativ errors  $\epsilon_{xi}, \boldsymbol{\epsilon}_{0i}, \boldsymbol{\epsilon}_{1i}$ are independent,
 the
augmented likelihood including the unobserved latent utilities is given as
\begin{align*} p(\xv,\xv^*,\yv|\Thetav,\fv)
=\prod_{i=1}^n p(x_i,x_i^*|\alphav,\lambda_x,f_{ci}) & \cdot
\prod_{i: x_i=0}
p(\yv_{0i}|\betav,\mathbf{S}_{0},\lambdav_{0},\zetav_0, f_{ci}, f_{0i} ) \cdot \\
  & \cdot \prod_{i: x_i=1}
p(\yv_{1i}|\betav,\mathbf{S}_{1},\lambdav_{1},\zetav_1, f_{ci},  f_{1i}),
\end{align*}
Here  $\fv=(\fv_c,\fv_0,\fv_1)$, where
$\fv_c$, $\fv_0$ and $\fv_1$ denote the vectors of latent factors for all
subjects and $\Thetav$ subsumes the regression effects $\alphav$ and $\betav$,
the factor loadings $\lambda_x, \lambdav_0, \lambdav_1, \zetav_0, \zetav_1$ and
the error variances $\sigmav^2_0=\Diag{\mathbf{S}_{0}}$ and
 $\sigmav^2_1= \Diag{\mathbf{S}_{1}}$.

Conditional on the latent factors $\fv$,
the models for the latent utilities and the potential outcomes
 are regression models with the respective  factors as additional regressors. This suggests to
 sample $(\alphav,\lambda_x)$ as well as  $(\betav,\lambdav_0, \lambdav_1, \zetav_0, \zetav_1)$ in
 one  block given $\fv$ and hence to
use an MCMC scheme which comprises the following steps:
\begin{itemize}
	\item[(1)]
 For $j=0,1$, sample the idiosyncratic variances $\sigmav^2_j$
  from $p(\sigmav^2_j|\Thetav_{\setminus \sigmav^2_j}, \yv^j, \fv)$
where $\yv^j=\{\yv_{i}|x_i=j\}$.

\item[(2)] For $i=1,\dots,n$, depending on treatment ($x_i=1$) or  control ($x_i=0$),
 sample the common latent factor $f_{ci}$ jointly with the specific latent
  factor $f_{x_i,i}$
 from $p(f_{ci},f_{x_i,i}|\Thetav, x_i, x_i^*,\yv_{i})$.

\item [(3)] For $i=1,\dots,n$, sample the latent utility $x_i^*$ from
$p(x_i^*|\Thetav,f_{ci}, x_i)$.

\item [(4)] Perform variable selection in the selection equation
and sample the indicators $\deltav^{\alpha}$ of
the  coefficients in  $\alphav$ subject to variable selection,
the  corresponding unrestricted  coefficients in  $\alphav$
 and $\lambda_x$ from
$p(\deltav^{\alpha}, \alphav, \lambda_x|\Thetav _{\setminus \alphav},\fv_c,\xv^*)$.

\item [(5)] Perform variable selection in both potential outcome models
and sample the indicators $\deltav^{\beta}$ of
the  coefficients in  $\betav$ subject to variable selection,
the  corresponding unrestricted  coefficients in  $\betav$
and the factor loadings $\lambdav_0,
\lambdav_1,\zetav_0,\zetav_1$ from
$p(\deltav^{\beta}, \betav, \lambdav_0,\lambdav_1, \zetav_0, \zetav_1 |
\Thetav _{\setminus \betav} ,\fv, \yv^0, \yv^1 )$.

\item[(6)] Perform a boosting step and a sign-switch for the latent factors and the corresponding factor loadings.

\item[(7)] Sample the hyperparameters $\pi_\alpha$
and $\pi_\beta$  from their respective posteriors
$p(\pi_\alpha| \deltav^\alpha)$ and $p(\pi_\beta| \deltav^\beta)$.
\end{itemize}

As draws are highly autocorrelated we apply  boosting based on marginal data augmentation
 as described in \cite{fru-lop:spa}. The full
 MCMC scheme is given in detail in Appendix~\ref{app:mcmc}.

\section{Simulation Example} \label{sec:sim}

\begin{figure}[t!]\centering
	\includegraphics[width=5.6cm,height=4.5cm,clip]{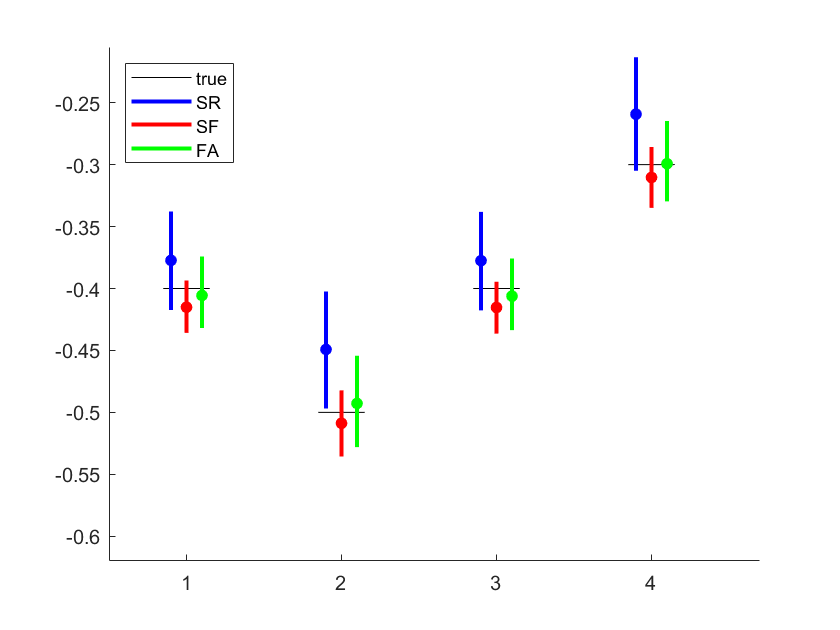} \hfill
	\includegraphics[width=5.6cm, height=4.5cm]{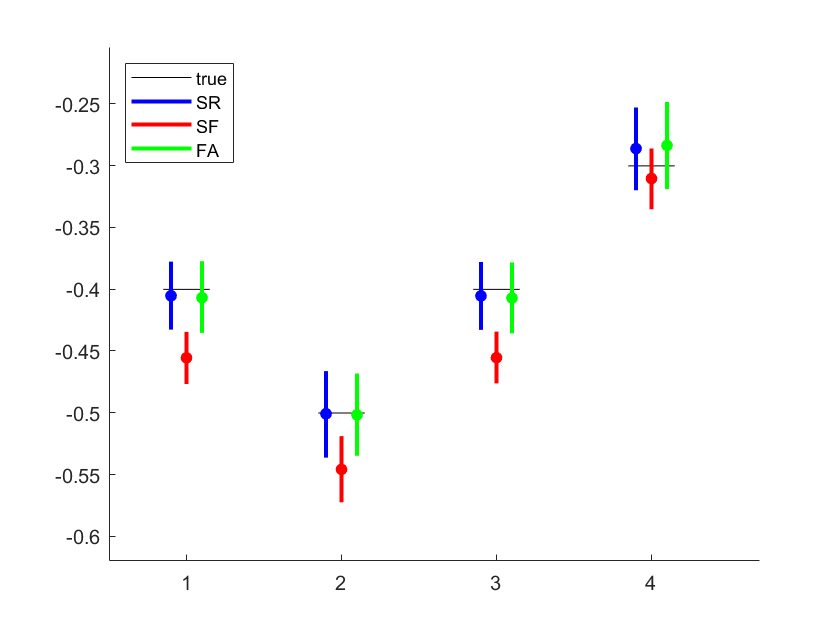}
		\caption{\label{wagner:fig1} True and estimated  in-sample average treatment effects
 (including 95\%-HPD intervals) for the various models. Data
		simulated from the  shared factor model (left) and the
switching regression model (right).}
\end{figure}

To illustrate the flexibility of the proposed FA treatment effects model, we
analyse two  data sets introduced in \cite{jac-etal:bay_tre}, which were simulated from the shared
 factor model (SF) and the switching regression model (SR), respectively,
  with parameters that violate  assumptions of the respective alternative model.  Each data set contains
   a panel of length $T=4$ of the observed outcome
of $n=50,000$ subjects where the large number of subjects  was chosen to illustrate the bias
in the average treatment estimates  resulting from miss-specification of the correlation structure.
For both data sets the probit model and the mean structure of the two potential outcomes models
 as well as the error variances $\sigmav_0$ and $\sigmav_1$ where specified to be identical
(see \cite{jac-etal:bay_tre} for details), the two data sets  however
 differed with respect to the correlation structure of the latent utility and the two potential outcomes sequences.

Both data sets  were analysed using a shared factor model, a switching regression model
 and a factor-augmented model (FA).
An estimate for the average treatment effects in panel  period $t$,
$\text{ATE}_t$,  over all subjects
  was obtained as $$\widehat{\text{ATE}}_t= \hat{\kappa}_t + \frac{1}{n} \sum_{i=1}^n
   \mathbf{w}_{it} \hat{\boldsymbol{\theta}},$$
   based on the estimated posterior means  of the  regression effects
   $\hat{\boldsymbol{\theta}}$ and  $\hat{\kappa}_t$ in the potential outcomes models.

Figure~\ref{wagner:fig1} shows  for $t=1,\dots,4$ the true average treatment effect and
the estimated  average treatment effects under each model  with the  95\%-HPD intervals.
 These intervals do not include  the true average treatment effect
at all time points if  data generated from the SF model are analysed with the SR model
or, vice-versa,
data generated from the SR model are analysed with the SF model. In contrast,
 in both cases the  proposed factor-augmented  model
 covers the true  treatment effects and performs similar as the respective data generating model.

\begin{figure}[t!]\centering
	\includegraphics[width=8cm, height=5cm,clip]{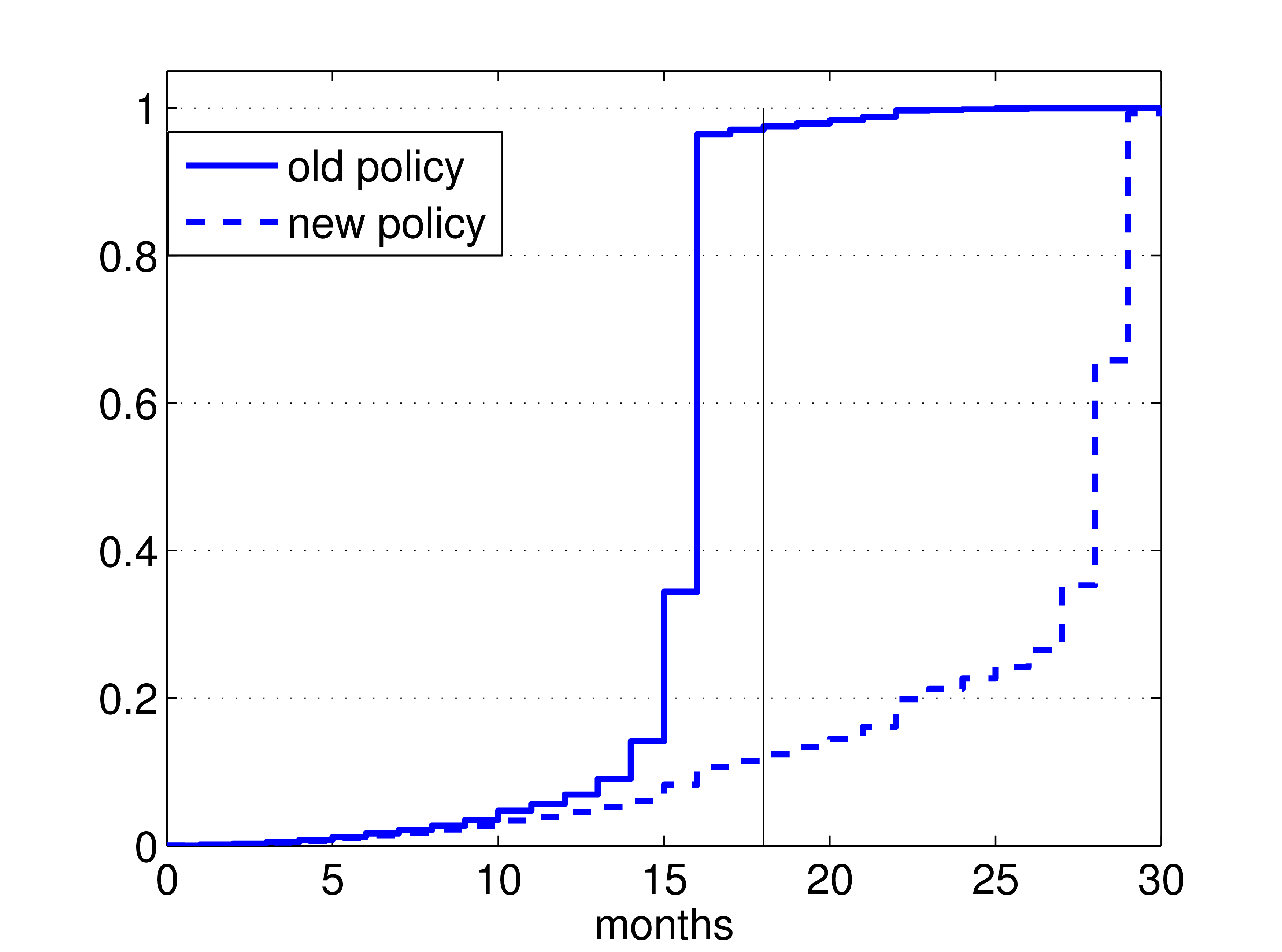} \hfill
		\caption{\label{wagner:fig2} Empirical cdf of the duration of maternity leave after child birth }
\end{figure}

\section{Analysing Earnings Effects of Maternity Leave} \label{sec:appl}

We apply the factor-augmented model to re-analyse the effects of a
 long  maternity leave  on earnings of Austrian mothers after their return to the labor market
  using the same  data as \cite{jac-etal:bay_tre}. The analysis is based  on  data  from the Austrian
  Social Security Data Base (ASSD), which is an administrative data set of the universe of Austrian
  employees providing detailed information on employment and maternity leave spells as well as demographic
   information on mothers  \citep{zwe-etal:aus}.

To exploit  a change in the parental leave policy in Austria in  July 2000 which extended the payment
of parental leave benefits  from 18 to 30 months, \cite{jac-etal:bay_tre} used data for mothers who
gave birth to  their last child  from June 1998 till July 2002. Figure \ref{wagner:fig2}
 illustrates that the majority of mothers returned to the labour market within 18  months
 before this policy change whereas
afterwards   most  mothers took a  longer maternity leave of more than 18 months.

\begin{figure}[t!]\centering
 \includegraphics[width=3.7cm, height=5.5cm,clip]{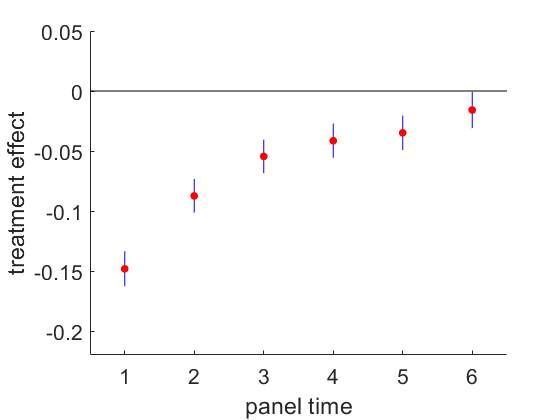} \hfill
     \includegraphics[width=3.7cm, height=5.5cm]{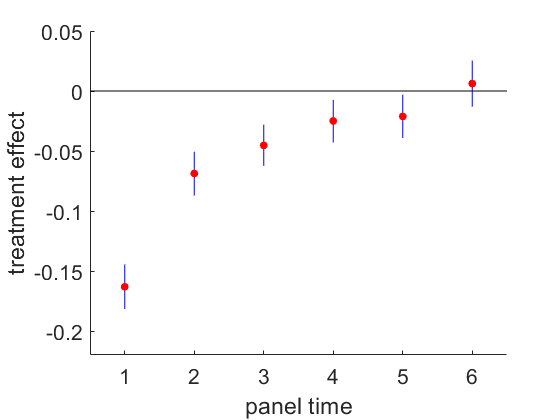}   \hfill
      \includegraphics[width=3.7cm, height=5.5cm]{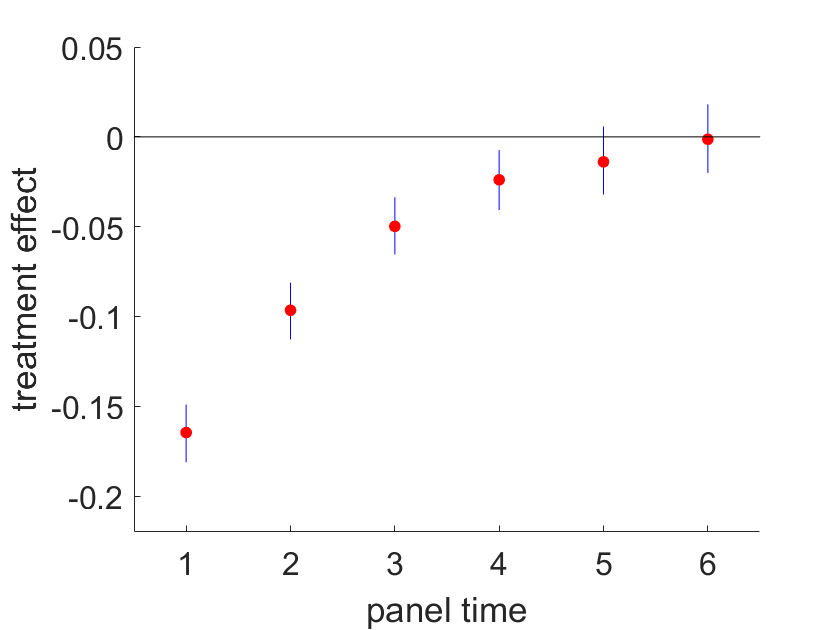}
\caption{\label{wagner:fig3} Estimated average treatment effects and 95\%-HPD-intervals of a
long maternity leave. Analysis with the  shared factor model (SF, left), the  switching
 regression model (SR, middle) and the factor-augmented model ({FA}, right).}
\end{figure}

 \cite{jac-etal:bay_tre} defined treatment as a maternity leave longer than 18 months.
   Based on a shared factor model as well as a switching regression model, they
    analysed the treatment effect on earnings  for those mothers who returned to the labour
    market immediately after the end of the maternity leave.
  We use the same specification for the mean of the latent utility
and the potential outcomes, defined as the log income,  as
in  their analysis, but   model the joint distribution of the errors
 $\boldsymbol{\varepsilon}_i$ by the  factor-augmented model proposed in this paper.

 Figure~\ref{wagner:fig3} compares the average treatment effects over all mothers
 in the sample  for the first 6 years after returning to the
 labour market, estimated by these  three  models. For all models,
  a  long maternity leave  results in considerably
lower earnings in the first panel period  with  the gap  decreasing over time. However, the evolvement
of  $\widehat{\text{ATE}}_t$ is slightly different for the three models: it is still negative
 in panel period 6 for the shared factor model, positive for the switching regression model
 and practically zero for the  factor-augmented model.
Detailed estimation results are provided in  Appendix~\ref{AppMother}
in Table~\ref{tab:selres}, Table~\ref{tab:outres}  and
 Table~\ref{tab:varres}.

\section{Conclusion} \label{sec:conclusion}
Inference on treatment effects for longitudinally observed outcomes  can be biased when  the
model used for data analyses implies restrictions on the  association between selection into
treatment and the potential outcomes sequences  as well as within the potential outcomes  sequences which are violated for the data to be analysed.
The proposed factor-augmented model  explicitly models
these associations by latent factors and  hence is more flexible than the models used so far.
However, as the two potential outcomes sequences are never observed together, only the fit of
the probit model for the latent utility and the
marginal models for the potential outcomes can be
assessed, but not the implied association of the two potential outcomes.

\bibliographystyle{plainnat}
\bibliography{sylvia_kyoto}

\begin{appendix}
	\section{Moments of  the observed outcomes}\label{app:obsmoments}
To derive the  first two moments of the observed outcomes we  start with  a univariate normal random variable $Z \sim \Normal{\mu,\sigma^2}$,
 and then consider the $(T+1)$-variate normal random variable $(x^*,\yv)'$.

	Expectation and variance  of $Z$ truncated to $(a,b)$ are given as
	\begin{align*}
		\E(Z|a<Z<b) &=\mu- \sigma\frac{\phi(\frac{b-\mu}{\sigma})-\phi(\frac{a-\mu}{\sigma})}
{\Phi(\frac{b-\mu}{\sigma})-\Phi(\frac{a-\mu}{\sigma})}, \\
		\V(Z|a<Z<b) &=\sigma^2 \left[
1- \frac{1}{\sigma}\frac{(b-\mu)\phi(\frac{b-\mu}{\sigma})-(a-\mu)\phi(\frac{a-\mu}{\sigma})}
{\Phi(\frac{b-\mu)}{\sigma}-\Phi(\frac{a-\mu}{\sigma})} -\Big(\frac{\phi(\frac{b-\mu}{\sigma})
-\phi(\frac{a-\mu}{\sigma})}{\Phi(\frac{b-\mu}{\sigma})-\Phi(\frac{a-\mu}{\sigma})}\Big) \right]^2.
	\end{align*}
	Let $(x^*,\yv)'\sim \Normal{\muv, \Sigmav}$
		with moments
	$$\muv=\begin{pmatrix} \mu_x \\ \muv_\yv \end{pmatrix} \quad \text{and} \quad
	\Sigmav= \begin{pmatrix} \sigma^2_x & \sigmav_{x \yv}'\\
	\sigmav_{x \yv} & \Sigmav_\yv \end{pmatrix}.$$
	The conditional distribution of $\yv|x^*$ is given as
	$$\yv|x^* \sim \Normal{\muv_\yv +\frac{\sigmav_{x \yv}}{\sigma_x^2} (x^*-\mu_x) ,
\Sigmav_{\yv}- \frac{\sigmav'_{x \yv}\sigmav_{x \yv}}{\sigma^2_x}},$$
 and interest is  in the first two moments of 	$\yv|(x^*<0)$ as well as  those of $\yv|(x^*>0)$.
 
The conditional expectation $\E(\yv|x^*<0)$ results
	as
	$$\E(\yv|x^*<0)  =\muv_\yv +\frac{\sigmav_{x \yv}}{\sigma_x^2}
\E(x^*-\mu_x|x^*<0) 
	=\muv_\yv -\frac{\sigmav_{x \yv}}{\sigma_x} \frac{\phi(\mu_x/\sigma_x)}{1-\Phi(\mu_x/\sigma_x)}.$$
The conditional second moment of  $(\yv-\muv_\yv)|(x^*<0) $ can be derived as
$$\E((\yv-\muv_\yv)(\yv-\muv_\yv)'|x^*<0)   =	
\Sigmav_{\yv} -\frac{\sigmav'_{x \yv}\sigmav_{x \yv}}{\sigma^2_x}+
 \frac{\sigmav'_{x \yv}\sigmav_{x \yv}}{\sigma^4_x} \E\big((x^*-\mu_x)^2|x^*<0),$$
	and, hence,
	  the conditional covariance matrix $\V(\yv |x^*<0)$ is given as
	  \begin{align*}\V(\yv |x^*<0)
	  &=	  	\Sigmav_{\yv} -	 \frac{\sigmav'_{x \yv}\sigmav_{x \yv}}{\sigma_x^2}
	 +  \frac{\sigmav'_{x \yv}\sigmav_{x \yv}}{\sigma_x^2}\V(x^*|x^*<0)\\
	&=\Sigmav_{\yv}+ \frac{\sigmav'_{x \yv}\sigmav_{x \yv}}{\sigma_x^2}
\Big(\frac{\mu_x}{\sigma_x}\frac{\phi(\mu_x/\sigma_x)}{1-\Phi(\mu_x/\sigma_x)}
	- \big(\frac{\phi(\mu_x/\sigma_x)}{1-\Phi(\mu_x/\sigma_x)})^2\Big).
	\end{align*}
	Similarly,  expectation and covariance of $\yv|(x^*>0)$ can be derived as
	 \begin{align*}
	 \E(\yv|x^*>0)&=\muv_\yv +\frac{\sigmav_{x \yv}}{\sigma_x}\frac{\phi(\mu_x/\sigma_x)}{\Phi(\mu_x/\sigma_x)}\\
	 \V(\yv |x^*>0) &=\Sigmav_{\yv} -\frac{\sigmav'_{x \yv}\sigmav_{x \yv}}{\sigma^2_x}\Big(\frac{\mu_x}{\sigma_x}\frac{\phi(\mu_x/\sigma_x)}{\Phi(\mu_x/\sigma_x)}
	 + \big(\frac{\phi(\mu_x/\sigma_x)}{\Phi(\mu_x/\sigma_x)})^2\Big).
	 	\end{align*}
 	\section{MCMC scheme}\label{app:mcmc}

 	With starting values for $\alphav, \betav, \zetav,\lambdav$
 and all latent factors MCMC is performed by iterating the following steps:
 	\begin{enumerate}
 		
 		\item[(1)] {\bf Sample the idiosyncratic variances $\sigmav^2_0$ and  $\sigmav^2_1$.}
 		For $j=0,1$ and $t=1,\dots, T$ sample  $\sigma_{jt}^2$ from
 		$\Gammainv{s_{n,jt},S_{n,jt}}$ where
 		$$ s_{n,jt}=s_{0,jt}+ n_{jt}/2, \qquad  S_{n,jt}=S_{0,jt}+ Se_{jt}/2,$$
 		and
 		$$ Se_{jt}=\sum_{i: x_i=j} (y_{j,it}-\wv_{j,it} \betav-f_{ci} \lambda_{jt} - f_{ji} \zeta_{j,t})^2.$$
 		Here $n_{jt}$ is  the number of subjects for which  $y_{j,it}$ is observed and
 $\wv_{j,it}$ denotes the values of the covariates at panel time $t$, i.e.
 		row  $t$ of the covariate  matrix $\Wv_{ji}$.

 		\item[(2)]  {\bf Sample the latent factors.}
 		For $i=1,\dots,n,$ sample the latent factor $f_{ci}$  and the specific factor $f_{ji}$ for $x_i=j$ from the full conditional posterior
 		$$p(f_{ci}, f_{ji}|\Thetav,x_i^*,\yv_{x_i,i})\propto p(x^*_i,\yv_{x_i,i}|\Thetav,f_{ci})p(f_{ci}, f_{ji})$$
 		
 		For $x_i=j$, the  errors of  latent
  utility  $x^*_i$ and the outcome vector $\yv_{ji}$  are  given as
 		\begin{align*} \varepsilon_{xi} &=x^*_i- \vv_i \alphav  = \lambda_x f_{ci} +\epsilon_{xi},\\
 			\varepsilonv_{ji} & = \yv_{ji}- \Wv_i \betav =\lambdav_j f_{ci} +
  \zetav_{j} f_{ji}+\epsilonv_{ji},
 		\end{align*}
 		and hence the full conditional of $(f_{ci}, f_{ji})$  is a bivariate normal  distribution,
 		$\Normal{\fv_{n,i},\Fv_{n,i}}$.
 		With $$\Psiv_j=\begin{pmatrix} \lambda_x & 0 \\ \lambdav_j & \zetav_j \end{pmatrix}
 \qquad \text{and} \qquad \tilde{\Sv_j}= \diag(\sigma^2_x, \sigmav_j ),$$ the posterior moments are given as 		
 		\begin{align*}
 			\Fv_{n,i}  & = ( \Psiv_j' \tilde{\Sv_j}^{-1}
 \Psiv_j +\Iv_2)^{-1},\\
 			\fv_{n,i} & = \Fv_{n,i} \Psiv_j' \tilde{\Sv_j}^{-1}
 \begin{pmatrix} \varepsilon_{xi}\\ \varepsilonv_{ji} \end{pmatrix}.
 		\end{align*}
 		 		
 		\item[(3)] {\bf Sample the latent utilities $x^*_i$.}
  		For $i=1,\dots,n$ sample  $x^*_i$ from $\Normal{\vv_i \alphav+\lambda_x f_{ci}, 1}$ truncated to the interval $(-\infty,0)$
 		for $x_i=0$ and to  $(0,\infty)$ if $x_i=1$.
 		
 		\item[(4)] {\bf  Sample the  parameters in the  selection equation.}
 The selection equation for $x_i^*$, $i=1,\dots, n$, is  given   as
   				$$x_i^*= \vv_i \alphav+ f_{ci} \lambda_x+ \epsilon_{xi},
    \quad  \epsilon_{xi} \sim \Normal{0,1}.$$
 		$f_{ci}$ is mandatorily included in the model and therefore
 		only elements of $\alphav$ are subject to  variable selection,  but $\lambda_x$ is not.

 		\item[(5)] {\bf Sample the  parameters of the outcome equation.} 		
 		The model for the observed outcomes $\yv_{x_i,i}$, $i=1,\dots, n$ is
  given   as
 		$$\yv_{x_i,i}=\Wv_{x_i,i} \betav + f_{ci} \lambdav_{x_i}  + f_{x_i,i} \zetav_{x_i}+
 \epsilonv_{x_i,i}, \quad \epsilonv_{x_i,i} \sim \Normal{\mathbf{0},\Sigmav_{x_i}}.$$
 		Also in the outcome equation,
 variable selection is  implemented  only for the elements of $\betav$, but
  not for the factor loadings
 		$\lambdav_{x_i}$ and  $\zetav_{x_i}$

 		\item[(6)]  {\bf Boosting.} Perform boosting based on marginal data augmentation as described
  in \cite{fru-lop:spa} and a sign-switch for the factor loadings and the respective factor.
 		For the sign-switch,   $\tau_c$, $\tau_0$, and $\tau_1$ are sampled
 		independently from $\{-1,1\}$ and
 		\begin{align*}
 		\lambdav&=\tau_c \lambdav,  \quad  f_{ci}=\tau_c \lambdav, \quad  i=1,\dots, n,\\
 		\lambdav&=\tau_j \zetav,  \quad  f_{ji}=\tau_j \lambdav, \quad  i=1,\dots, n.
 		\end{align*}
 		
 		\item[(7)] {\bf Sample the inclusion probabilities.}
       Sample  $\pi_\alpha | k_{\alpha}$ from  $\Betadis{1+k_{\alpha},1+d_{\alpha}-k_{\alpha}}$
         and $\pi_\beta | k_{\beta}$ from
         $\Betadis{1+k_{\beta},1+d_{\beta}-k_{\beta}}$
 		where $k_{\alpha}=\sum \delta^\alpha_{\ell}$ is the number of selected
 regressors for the latent utility  and  $k_{\beta}=\sum \delta^\beta_{\ell}$
 accordingly the number of selected
 		regressors for the potential outcome equations.
 		
 	\end{enumerate} 	
 	
 	Sampling steps (4) and (5) are  standard sampling steps in linear regression models with variable selection,
 see \cite{jac-etal:bay_tre} for details.
 	
 \begin{table}[t!]
\begin{center}
\begin{tabular}{|l|lll|}
	\hline
 	      & mean & sd   & prob  \\ \hline
intercept & -1.562   &0.032 & --- \\
z         &  2.827   &0.021 &  1.000 \\
child 2   &  0.052   &0.036 &  0.737 \\
child $>3$& -0.012   &0.032& 0.157 \\
exp       &  0.093   &0.027 & 0.985 \\
blue collar& -0.059  &0.044 & 0.706 \\
int. exp/blue & -0.016   & 0.040 & 0.177 \\
base-earn Q2  & 0.002   & 0.012  & 0.060 \\
base-earn Q3   & -0.001   & 0.008 & 0.035 \\
base-earn Q4   & -0.153   & 0.027 & 1.000 \\
\hline		
\end{tabular}
\caption{Results selection equation: posterior means (mean), standard deviation (sd, in parentheses)
and estimated posterior inclusion probabilities (prob.) of standardized regression effects}
\label{tab:selres}
\end{center}
\end{table}
	
 \section{Results of the FA model for the mother data} \label{AppMother}

 The data set  contains information on 31,051 mothers
   with earnings  after return to labor market observed over  4-6 consecutive panel periods.
Covariates are specified  as in the analysis of \cite{jac-etal:bay_tre}.
Covariates included in
the selection equation are an indicator for the policy change (\emph{z}=1 indicates longer
 payment of parental leave benefits), indicator variables for
the child (\emph{child2}=1 if the mother has already an child  and
 \emph{child $>3$} =1 if the mother has 2 or more older children),
 the working experience (\emph{exp}=1 if the working experience is above the median working experience
 in the sample), type of contract (\emph{blue collar} or white collar)
and the interaction between these two and finally 
indicators to control for  earnings before the first child  in terms of quartiles.
The outcome model additionally includes indicator variables for  panel periods 2-6 and for return to the same employer (\emph{eq.emp}) and a quadratic calendar year effect.

 MCMC estimation as outlined in Appendix~\ref{app:mcmc}
 was run for 500000 iteration after a burn-in of 10000 with
 variable selection starting after 5000 iterations of the burn-in. To determine
  posterior mean estimates we applied a thinning of 500.

\begin{table}[t!]
\begin{center}
\begin{tabular}{|l|cc|cc|}
	\hline
& \multicolumn{2}{c|}{treatment 0} &\multicolumn{2}{c|}{treatment 1}\\
	 & mean \; (sd)  & prob  & mean \;  (sd)  & prob \\ \hline
intercept      & 9.309  \; (0.014) & ---  & -0.130   \; ( 0.011) & 1.000 \\
child 2        & -0.000 \; (0.001) & 0.008 & -0.000   \; ( 0.001) & 0.011 \\
child $>$3     & 0.000  \; (0.001) & 0.014 & 0.000   \;  (0.002)  & 0.021  \\
exp            & -0.092 \; (0.010) & 1.000 & 0.011   \; ( 0.015)  & 0.388 \\
blue collar    & -0.108 \; (0.006) & 1.000 & 0.000   \; ( 0.002)  & 0.015 \\
int. exp/blue  & 0.001  \; (0.004) & 0.039 & 0.005   \; ( 0.012)  & 0.173 \\
base-earn Q2   & 0.066  \; (0.006) & 1.000 & 0.000   \; ( 0.002)  & 0.016 \\
base-earn Q3   & 0.286  \; (0.011) & 1.000 & -0.047   \; ( 0.014) & 0.974 \\
base-earn Q4   & 0.606  \; (0.010) & 1.000 & -0.116   \; ( 0.013)  & 1.000 \\
eq. emp        & 0.049  \; (0.005) & 1.000 & 0.000   \; ( 0.003)  & 0.031 \\
panel $t=2$    & 0.066  \; (0.004) & 1.000 & 0.068   \; ( 0.004)  & 1.000 \\
panel $t=3$    & 0.106  \; (0.006) & 1.000 & 0.115   \; ( 0.006)  & 1.000 \\
panel $t=4$    & 0.149  \; (0.009) & 1.000 & 0.141   \; ( 0.008)  & 1.000 \\
panel $t=5$    & 0.201  \; (0.011) & 1.000 & 0.151  \; ( 0.009)   & 1.000 \\
panel $t=6$    & 0.252  \; (0.013) & 1.000 & 0.163   \; ( 0.010   & 1.000 \\ \hline
$(year-1999)$     & \multicolumn{4}{c|}{ 0.050   \;  (0.005)} \\
$(year-1999)^2 $   &\multicolumn{4}{c|}{-0.005  \;  (0.000)}   \\ \hline	
 \end{tabular}			
\caption{Results outcome equation: posterior means (mean), standard deviation (sd, in parentheses)
and estimated posterior inclusion probabilities (prob.) of  regression effects}
\label{tab:outres}
\end{center}
\end{table}

Table~\ref{tab:selres} reports results for the
standardized estimated effects $\tilde{\alphav}$.
The factor loading of the common factor in the
selection equation is  $\hat{\lambda}_x= 0.277$ (sd=0.014).
 Estimation results for the regression effects in the outcome equation   are
 given in Table~\ref{tab:outres}  and  for the factor loadings and idiosyncratic variances
  in Table~\ref{tab:varres}.

 \begin{table}[t!]
 \begin{center}
 \begin{tabular}{|l|ccc|ccc|}\hline
 & \multicolumn{3}{c|}{treatment 0} &\multicolumn{3}{c|}{treatment 1}\\	
t & $\lambdav_0$ & $\zetav_0$ & $\sigmav_0^2$& $\lambdav_1$ & $\zetav_1$ & $\sigmav_1^2$\\\hline
1 & -0.296 (0.017)  &0.254 (0.019)&0.088 (0.001) & 0.327 (0.005)& 0.038 (0.028)  &  0.078 (0.001) \\
2 & -0.334 (0.019)  & 0.284 (0.022)&0.025 (0.001) & 0.385 (0.009) &0.068 (0.039) &0.016 (0.001) \\
3 & -0.377 (0.012)  & 0.183 (0.024)& 0.038 (0.001) & 0.340 (0.020) &  0.171 (0.038) & 0.023 (0.000) \\
4 & -0.412 (0.006)  & 0.083 (0.026)& 0.032 (0.000)& 0.288 (0.031) &0.271 (0.032) & 0.019 (0.000) \\
5 & -0.434 (0.003)  & 0.023 (0.016)& 0.012 (0.000) & 0.248 (0.035) &  0.310 (0.027) & 0.023 (0.001) \\
6 & -0.415 (0.003)  & 0.022 (0.016)& 0.032 (0.001) & 0.231 (0.035) &  0.308 (0.026) &0.037 (0.001)  \\    \hline
\end{tabular} 	 			
\caption{Results factor loadings and idiosyncratic variances: posterior means (mean) and standard deviations (sd, in parentheses)}
\label{tab:varres}
\end{center}
\end{table}
 		
	\end{appendix}
\end{document}